\RequirePackage{fix-cm}
\documentclass[a4paper,11pt]{article}

\usepackage{authblk}
\usepackage[utf8]{inputenc}
\usepackage[T1]{fontenc}
\usepackage{pslatex}
\usepackage{amsfonts,amsmath,amssymb,amsthm}
\usepackage{algorithm}
\usepackage{paralist}
\usepackage{booktabs}
\usepackage{url}
\usepackage{hyperref}
\usepackage{graphicx}
\usepackage{microtype}
\usepackage{url}
\usepackage{hyperref}
\usepackage{makecell}
\usepackage{subcaption}
\usepackage{tikz}
\usepackage{multirow}
\usetikzlibrary{positioning}
\usetikzlibrary{decorations.pathreplacing}
\newcommand{\N}{\mathbb{N}}

\newcommand{\F}{\mathbb{F}}

\newtheorem{definition}{Definition}
\newtheorem{lemma}{Lemma}
\newtheorem{theorem}{Theorem}

\newtheorem{problem}{Problem}
\newtheorem{corollary}{Corollary}

\tikzset{
    mybrace/.style={decorate,decoration={brace,aspect=#1}}
}

\providecommand{\keywords}[1]{\textbf{\textit{Keywords }} #1}

\begin{document}

\title{Self-Orthogonal Cellular Automata}

\author[1]{Luca Mariot}
\author[1]{Federico Mazzone}
	
\affil[1]{{\small Semantics, Cybersecurity and Services Group, University of Twente, Drienerlolaan 5, 7511GG Enschede, The Netherlands} 
	
	{\small \texttt{\{l.mariot, f.mazzone\}@utwente.nl}}}

\maketitle

\begin{abstract}
It is known that no-boundary Cellular Automata (CA) defined by bipermutive local rules give rise to Latin squares. In this paper, we study under which conditions the Latin square generated by a bipermutive CA is self-orthogonal, i.e. orthogonal to its transpose. We first enumerate all bipermutive CA over the binary alphabet up to diameter $d=6$, remarking that only some linear rules give rise to self-orthogonal Latin squares. We then give a full theoretical characterization of self-orthogonal linear CA, by considering the square matrix obtained by stacking the transition matrices of the CA and of its transpose, and determining when it is invertible. Interestingly, the stacked matrix turns out to have a circulant structure, for which there exists an extensive body of results to characterize its invertibility. Further, for the case of the binary alphabet we prove that irreducibility is a sufficient condition for self-orthogonality, and we derive a simpler characterization which boils down to computing the parity of the central coefficients of the local rule.
\end{abstract}


\section{Introduction}
\label{sec:intro}
Cellular Automata (CA) provide an interesting framework for the construction of cryptographic primitives, including pseudorandom number generators~\cite{wolfram85,formenti14,leporati14}, S-boxes~\cite{picek17,ghoshal18,mariot19}, and secret sharing schemes~\cite{rey05,mariot14,herranz18}. The case of secret sharing schemes is particularly interesting for the type of combinatorial structures that arise from the one-shot behavior of CA, which contrasts with the usual long-term and asymptotic approach employed in the literature to study the dynamical properties of CA. Among the set of works that studied the combinatorial structures generated by CA in the context of secret sharing, the research thread focused on Latin squares proved to be quite fruitful in the last few years. In particular, it is known that the Cayley table of a \emph{bipermutive} CA (i.e. CA that are permutive in both the leftmost and rightmost coordinates of the local rule) is a Latin square, and that two bipermutive CA defined by linear local rules are orthogonal if and only if their associated polynomials are coprime~\cite{mariot20}. This result gives a straightfoward way to implement threshold secret sharing schemes where only two players are required in order to reconstruct the secret. Subsequent works explored different perspectives of this characterization of orthogonal CA, such as its specialization over more particular classes of Latin squares~\cite{hammer23}, the construction of Boolean functions with significant cryptographic properties~\cite{gadouleau23}, and the design subspace codes with optimal minimum distance~\cite{mariot23}.

In this paper, we take a different direction and consider the following problem: \emph{when is the Latin square generated by a bipermutive CA self-orthogonal, i.e. orthogonal to its transpose?} Beside their own theoretical interest, self-orthogonal Latin squares are connected to several other types of combinatorial designs, including \emph{resolvable Balanced Incomplete Block Designs} (BIBDs)~\cite{baker83}. In turn, resolvable BIBDs can be used to design anonymous secret sharing schemes~\cite{blundo97} and Mutually Unbiased Bases (MUBs) for quantum error-correcting codes~\cite{kumar22}.

We start our investigation of self-orthogonal CA by first exhaustively enumerating all bipermutive CA over the binary alphabet up to diameter $d=6$, and notice that among them only a few linear ones generate self-orthogonal Latin squares. We then provide a full theoretical characterization of self-orthogonal linear CA. In particular, we consider the square matrix obtained by stacking the transition matrix of the CA with that of its transpose, and studying under which conditions this matrix is invertible. Interestingly, we show that this stacked matrix is circulant, a remark that allows us to reduce the problem of its invertibility to the computation of the greatest common divisor of the polynomial associated with the local rule of the CA and the polynomial $X^{2(d-1)} - 1$. Additionally, we focus on the binary case where we prove that irreducibility of the polynomial associated to the local rule is a sufficient condition for self-orthogonality, and we derive a simpler characterization when the size of the stacked matrix is a power of 2. In particular, this characterization reduces the problem to determining the parity of the central coefficients of the linear rule.

The rest of this paper is structured as follows. Section~\ref{sec:bg} covers the basic definitions and results about cellular automata and Latin squares used throughout the paper. Section~\ref{sec:emp} formally states the problem of finding self-orthogonal CA, and discuss the preliminary results obtained from the exhaustive search experiments of bipermutive CA up to diameter $d=6$. Then, Section~\ref{sec:lin-ca} formally analyzes the case where the underlying local rule is linear, deriving a general characterization result for linear self-orthogonal CA over a finite field, and providing more refined results for the binary case. Finally, Section~\ref{sec:outro} recaps the main contributions of the paper, and describes a few interesting directions for future research on the topic.

\section{Background}
\label{sec:bg}
This section recalls the background terminology and results that the rest of the paper is based upon. In particular, the section covers first the basic concepts and representations concerning cellular automata, and then it reviews the essential results related to combinatorial designs and orthogonal Latin squares.

\subsection{Cellular Automata}
\label{subsec:ca}
A Cellular Automaton (CA) can be broadly conceived as a computational model defined by a regular lattice of cells. Commonly, this lattice is one-dimensional, although generalizations to more dimensions are possible. Each cell updates its state by evaluating a local rule on its neighborhood, which comprises the cell itself and some of its neighbors. A single step of computation amounts to applying in parallel the local rule at all cells.

Different settings can arise depending on how one addresses the update of the cells at the boundaries of the lattice. In this paper, we consider the \emph{no boundary} and \emph{periodic boundary} conditions, which lead us to the following definition:

\begin{definition}
	\label{def:ca}
	Let $\Sigma$ be a finite set of size $q \in \N$, and let $d, n \in \N$ such that $d\le n$. Further, let $f: \Sigma^d \to \Sigma$ be a $d$-variable mapping over $\Sigma$. Then, we define the following two models
	of \emph{one-dimensional} CA over the alphabet $\Sigma$ and with $n$ input cells, equipped with the \emph{local rule} $f$ of \emph{diameter} $d$:
	\begin{itemize}
		\item \emph{No Boundary CA} (NBCA): $F: \Sigma^{n} \rightarrow \Sigma^{n-d+1}$ is
		defined for all $x \in \Sigma^n$ as:
		\begin{equation}
		\label{eq:nbca}
		F(x_1, \ldots, x_n) = (f(x_1, \ldots, x_d), f(x_2, \ldots, x_{d+1}),
		\ldots, f(x_{n-d+1}, \ldots, x_n)) \enspace .
		\end{equation}
		\item \emph{Periodic Boundary CA} (PBCA): $F: \Sigma^{n} \rightarrow \Sigma^{n}$ is
		defined for all $x \in \Sigma^n$ as:
		\begin{equation}
		\label{eq:pbca}
		F(x_1, \ldots, x_n) = (f(x_1, \ldots, x_d), \ldots,
		f(x_{n-d},\ldots,x_1), \ldots, f(x_n,\ldots, x_{d-1}) \enspace .
		\end{equation}
	\end{itemize}  
\end{definition}
In general, the $i$-th output coordinate of a one-dimensional CA is determined by applying the local rule $f$ to the neighborhood formed by the $i$-th input cells and the $d-1$ cells to its right. The difference between the NBCA and PBCA model is that in the former the local rule is applied only to those cells that have enough neighbours on the right, while in the latter it is applied to all cells by considering the lattice as a ring modulo $n$.

The local map $f: \Sigma^d \to \Sigma$ is usually represented as a lookup table of $q^d$ entries which specifies the output value $f(x)$ for each input vector $x \in \Sigma^d$. In this context, the most basic setting corresponds to \emph{binary} CA, where the alphabet is the finite field of two elements $\Sigma = \F_2 = \{0, 1\}$. In particular, sum and multiplication in $\F_2$ corresponds respectively to the XOR (denoted as $\oplus$) and the logical AND (denoted by concatenation of the operands). A binary local rule can then be interpreted as a \emph{Boolean function} $f: \F_2^d \to \F_2$ of $d$ variables, where the input space is the $d$-dimensional vector space $\F_2^d$.

From a representation standpoint, the lookup table of $f$ is the \emph{truth table} representation of the Boolean function, and it can be uniquely identified by the $2^d$-bit string $\Omega_f \in \F_2^{2^d}$ of its outputs, once the input vectors of $\F_2^d$ have been sorted in lexicographic order. In the CA literature, it is customary to refer to the \emph{Wolfram code} of the local rule, which is simply the decimal encoding of $\Omega_f$~\cite{wolfram83}.

\begin{figure}[t]
    \centering
    \begin{subfigure}{0.5\textwidth}
        \centering
        \begin{tikzpicture}
            [->,auto,node distance=1.5cm, empt node/.style={font=\sffamily,inner
                sep=0pt}, rect
            node/.style={rectangle,draw,font=\bfseries,minimum size=0.5cm, inner
                sep=0pt, outer sep=0pt}]
            
            \node [empt node] (c)   {};
            \node [rect node] (c1) [right=0.1cm of c] {$1$};
            \node [rect node] (c2) [right=0cm of c1] {$0$};
            \node [rect node] (c3) [right=0cm of c2] {$0$};
            \node [rect node] (c4) [right=0cm of c3] {$1$};
            
            \node [empt node] (f1) [above=0.4cm of c2.east, xshift=0.25cm] {{\footnotesize
                    $f(1,0,0) = 1$}};
            
            \node [rect node] (p2) [above=0.85cm of c1] {$0$};
            \node [rect node] (p1) [left=0cm of p2] {$1$};
            \node [rect node] (p3) [right=0cm of p2] {$0$};
            \node [rect node] (p4) [right=0cm of p3] {$0$};
            \node [rect node] (p5) [right=0cm of p4] {$0$};
            \node [rect node] (p6) [right=0cm of p5] {$1$};
            
            \node [empt node] (p7) [below=0.2cm of p1] {};
            \node [empt node] (p8) [right=0.07cm of p7] {};
            \node [empt node] (p12) [above=0.5cm of p1.east] {};
            \node [empt node] (p13) [above=0.5cm of p5.east] {};
            \node [empt node] (p14) [above=0.3cm of p13] {\phantom{M}};
            
            \draw [-, mybrace=0.25, decorate, decoration={brace,mirror,amplitude=5pt,raise=0.3cm}]
            (p1.west) -- (p3.east) node [midway,yshift=-0.3cm] {};
            \draw [-, draw=white, decorate, decoration={brace,amplitude=5pt,raise=0.3cm}]
            (p1.west) -- (p2.east) node [midway,yshift=0.3cm] {};
            \draw[->] (p8) -- (c1.north);
            \draw[->, draw=white] (p12) edge[bend left] (p13);
        \end{tikzpicture}
        \caption{NBCA}
    \end{subfigure}%
    \begin{subfigure}{0.5\textwidth}
        \centering
        \begin{tikzpicture}
            [->,auto,node distance=1.5cm, empt node/.style={font=\sffamily,inner
                sep=0pt}, rect
            node/.style={rectangle,draw,font=\sffamily\bfseries,minimum size=0.5cm, inner
                sep=0pt, outer sep=0pt}, grey node/.style={rectangle,draw,fill=gray!40,
                font=\sffamily\bfseries,minimum size=0.5cm, inner sep=0pt, outer sep=0pt}]
            
            \node [empt node] (c)   {};
            \node [rect node] (c1) [right=0.1cm of c] {$0$};
            \node [rect node] (c0) [left=0cm of c1] {$1$};
            \node [rect node] (c2) [right=0cm of c1] {$0$};
            \node [rect node] (c3) [right=0cm of c2] {$1$};
            \node [rect node] (c4) [right=0cm of c3] {$0$};
            \node [rect node] (c5) [right=0cm of c4] {$0$};
            
            \node [empt node] (f1) [above=0.2cm of c3] {{\footnotesize
                    $f(1,1,0) = 0$}};
            
            \node [rect node] (p2) [above=0.85cm of c1] {$0$};
            \node [rect node] (p1) [left=0cm of p2] {$1$};
            \node [empt node] (p)  [left=0.1cm of p1] {};
            \node [rect node] (p3) [right=0cm of p2] {$0$};
            \node [rect node] (p4) [right=0cm of p3] {$0$};
            \node [rect node] (p5) [right=0cm of p4] {$0$};
            \node [rect node] (p6) [right=0cm of p5] {$1$};
            \node [grey node] (p7) [right=0cm of p6] {$1$};
            \node [grey node] (p8) [right=0cm of p7] {$0$};
            \node [empt node] (p9) [right=0.1cm of p8] {};
            
            \node [empt node] (p10) [below=0.2cm of p6] {};
            \node [empt node] (p11) [right=0.07cm of p10] {};
            \node [empt node] (p12) [above=0.5cm of p1.east] {};
            \node [empt node] (p13) [above=0.5cm of p7.east] {};
            
            \draw [-, mybrace=0.25, decorate, decoration={brace,mirror,amplitude=5pt,raise=0.3cm}]
            (p6.west) -- (p8.east) node [midway,yshift=-0.3cm] {};
            \draw [-, decorate, decoration={brace,amplitude=5pt,raise=0.3cm}]
            (p1.west) -- (p2.east) node [midway,yshift=0.3cm] {};
            \draw [-, decorate, decoration={brace,amplitude=5pt,raise=0.3cm}]
            (p7.west) -- (p8.east) node [midway,yshift=0.3cm] {};
            \draw[->] (p11) -- (c5.north);
            \draw[->] (p12) edge[bend left] (p13);
        \end{tikzpicture}
        \caption{PBCA}
    \end{subfigure}
    \caption{Examples of NBCA and PBCA evaluations under  local rule 150.}
    \label{fig:nbca-pbca}
\end{figure}
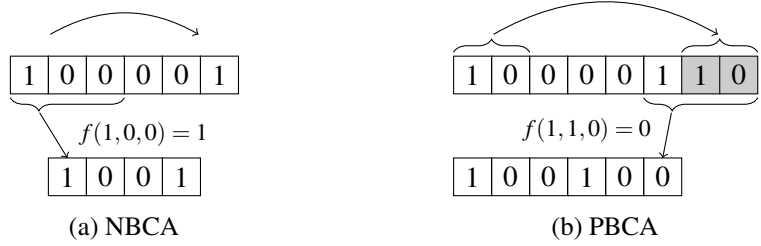

As an example, Figure~\ref{fig:nbca-pbca} displays an example of binary NBCA and PBCA of length $n=6$ equipped with the linear local rule 150, which is defined for all $x=(x_1,x_2,x_3) \in \F_2^3$ as $f(x_1, x_2, x_3) = x_1 \oplus x_2 \oplus x_3$.

Another method to uniquely represent Boolean functions that we will use throughout the paper for the local rules of binary CA is the \emph{Algebraic Normal Form} (ANF)~\cite{carlet21}. Formally, the ANF of $f: \F_2^d \to \F_2$ is the multivariate polynomial $P_f$ over the quotient ring $\F_2[x]/(x_1^2 \oplus x_1, \cdots, x_n^2 \oplus x_n)$ defined as follows:
\begin{equation}
\label{eq:anf}
P_f(x) = \bigoplus_{u \in \F_2^n} a_u x^u = \bigoplus_{u \in \F_2^n} a_u (x_1^{u_1}\cdots x_n^{u_n}) \enspace ,
\end{equation}
where $a_u \in \F_2$ for all $u \in \F_2^n$. The \emph{algebraic degree} of $f$ is the size of the largest nonzero monomial in $P_f$. Functions of degree 1 are also called \emph{affine}, and an affine function is \emph{linear} if the coefficient of its constant term $a_{\underbar{0}}$ is null.

Beside the binary case, in this paper we consider CA defined over the finite field $\F_q$ as an alphabet, mainly due the interest for applications in cryptography and coding theory~\cite{lidl94}. In this case, the size $q$ of the alphabet is a power of a prime number. Linear CA can be generalized also over the alphabet $\F_q$. Specifically, the function $f: \F_q^d \to \F_q$ is called linear if it is possible to write it as a linear combination of the input variables, that is:
\begin{equation}
    \label{eq:lin-rule}
    f(x_1,\ldots,x_d) = a_1x_1 + \ldots + a_dx_d 
\end{equation}
for all $x = (x_1,\ldots,x_d) \in \F_q^d$, where $a_1,\ldots a_d \in \F_q$. The global map of a CA $F: \F_q^n \to \F_q^{n-d+1}$ defined by the linear local rule $f$ in Eq.~\eqref{eq:lin-rule} can be expressed as the linear transformation $F(x) = M_F\cdot x^\top$, where $M_F$ is the $(n-d+1)\times n$ \emph{transition matrix} defined as: 
\begin{equation}
    \label{eq:mat-lin-ca}
    M_F = \begin{pmatrix}
        a_1    & \ldots & a_{d-1} & a_d & 0 & \ldots & \ldots & \ldots & \ldots & 0 \\
        0      & a_1    & \ldots  & a_{d-1} & a_d & 0 & \ldots & \ldots & \ldots & 0 \\
        \vdots & \vdots & \vdots & \ddots  & \vdots & \vdots & \vdots & \ddots & \vdots & \vdots \\
        0 & \ldots & \ldots & \ldots & \ldots & 0 & a_1 & \ldots & a_{d-1} & a_d \\
    \end{pmatrix} \enspace .
\end{equation}
Further, one can naturally associate to the linear local rule $f$ of Eq.~\eqref{eq:lin-rule} the polynomial $p_f \in \F_q[X]$ with the following form:
\begin{equation}
    \label{eq:lin-pol}
    p_f(X) = a_1 + a_2X + \ldots + a_{d}X^{d-1} \enspace .
\end{equation}
In other words, we assign the coefficients $a_1, \ldots, a_d \in \F_q$ to the increasing powers of the unknown $X$, thus obtaining a polynomial of degree at most $d-1$. From a coding-theoretic standpoint, $M_F$ and $p_f$ are respectively the generator matrix and the generator polynomial of a cyclic code~\cite{mariot18}.

Due to their connections with combinatorial designs and Latin squares, in this paper we also consider the specific class of \emph{permutive} CA. Formally, a local rule $f: \Sigma^d \to \Sigma$ is permutive in the $i$-th coordinate (with $i \in \{1\,\ldots, d\}$) if, by fixing all input other coordinates $j \neq i$ to a constant value, the resulting restriction of $f$ is a permutation over $\Sigma$. In the binary case $\Sigma = \F_2$, a permutive local rule $f: \F_2^d \to \F_2$ can be written as:
\begin{equation}
    \label{eq:iperm}
    f(x_1,\ldots, x_d) = x_i \oplus g(x_1,\ldots, x_{i-1}, x_{i+1},\ldots, x_i) \enspace ,
\end{equation}
where $g: \F_2^{d-1} \to \F_2$ is a function of $d-1$ variables, also called the \emph{generating function} of $f$~\cite{leporati14}. 

A function $f$ that is permutive in the first (respectively, the last) coordinate is also called \emph{leftmost} (respectively, \emph{rightmost}) \emph{permutive}. If $f$ is both leftmost and rightmost permutive, then it is called \emph{bipermutive}. The aforementioned rule 150 is an example of bipermutive rule. Clearly, any linear rule over $\F_q^d$ of the form defined in Eq.~\eqref{eq:lin-rule} is bipermutive if and only if both $a_1$ and $a_d$ are not null.

\subsection{Latin Squares}
\label{subsec:ls}
In this section we cover only the basic facts related to Latin squares, referring the reader to~\cite{denes-ls,stinson-cd} for a more thorough treatment of the subject. In what follows, we denote by $[n] = \{1,\ldots, n\}$ the set of the first $n \in \N$ integer positive numbers. We start by first giving the definition of Latin square:

\begin{definition}
\label{def:ls}
A Latin square of order $N \in \N$ is a $N \times N$ matrix $L$ with entries from $[N]$ such that for all $i, j, k \in [N]$ with $j\neq k$, one has that $L(i, j) \neq L(i, k)$ and $L(j, i) \neq L(k, i)$.
\end{definition}
Intuitively, Definition~\ref{def:ls} states that every number from $1$ to $N$ occurs exactly once in each row and in each column of a Latin square. Alternatively, each row and each column of $L$ forms a permutation of $[N]$.

The next definition introduces orthogonal Latin squares:
\begin{definition}
\label{def:ols}
Two Latin squares $L_1, L_2$ of order $N \in \N$ are orthogonal if for all distinct pairs of coordinates $(i, j), (i', j') \in [N] \times [N]$ it holds that
\begin{equation}
(L_1(i,j), L_2(i,j)) \neq (L_1(i', j'), L_2(i', j')) \enspace .
\end{equation}
\end{definition}
Thus, two Latin squares of the same order are orthogonal if their \emph{superposition} yields all ordered pairs of the Cartesian product $[N] \times [N]$ exactly once.

We now briefly review the construction of orthogonal Latin squares based on CA proposed in~\cite{mariot20}. To this end, we first introduce the \emph{Cayley table} associated to a generic NBCA, which requires a bit more of notation on how we can encode $(d-1)$-cell blocks as integer numbers. Specifically, we assume that $\Sigma^{d-1}$ is totally ordered, and that $\phi: \Sigma^{d-1} \rightarrow [N]$ is a
monotone bijective mapping between $\Sigma^{d-1}$ and $[N]$, where the latter inherits the usual order of natural numbers. The inverse mapping of $\phi$ is denoted by $\psi = \phi^{-1}$. This notation leads to the following definition:
\begin{definition}
    \label{def:cayley}
    Let $F: \Sigma^{2(d-1)} \rightarrow \Sigma^{d-1}$ be a NBCA defined by the local rule $f: \Sigma^d \rightarrow \Sigma$ of diameter $d$ over an alphabet $\Sigma$ of $q$ symbols. Setting $N = q^{d-1}$, the \emph{Cayley table} associated to $F$ is the $N \times N$ matrix $C_F$ with entries from $[N]$ such that
    \begin{equation}
        \label{eq:sq-ca}
        C_{F}(i,j) = \phi(F(\psi(i)||\psi(j))) \enspace ,
    \end{equation}
    for all $1 \le i,j \le N$, where $\psi(i)||\psi(j) \in \Sigma^{2(d-1)}$ denotes the \emph{concatenation} of  $\psi(i),\psi(j) \in \Sigma^{d-1}$.
\end{definition}
Intuitively, this means that the output of the CA corresponds to the entry of the Cayley table at the row and column coordinates respectively encoded by the left and the right half of the CA input configuration.

The next Lemma about CA and Latin squares has been independently rediscovered in multiple works under different guises~\cite{eloranta93,mariot20}:
\begin{lemma}
\label{lm:lat-sq-bip-ca}
Let $\Sigma$ be a finite alphabet of size $q$. The Cayley table $C_F$ of a CA $F: \Sigma^{2(d-1)} \rightarrow \Sigma^{d-1}$ defined by a bipermutive local
rule $f:\Sigma^{d}\rightarrow \Sigma$ is a Latin square of order $N=q^{d-1}$.
\end{lemma}

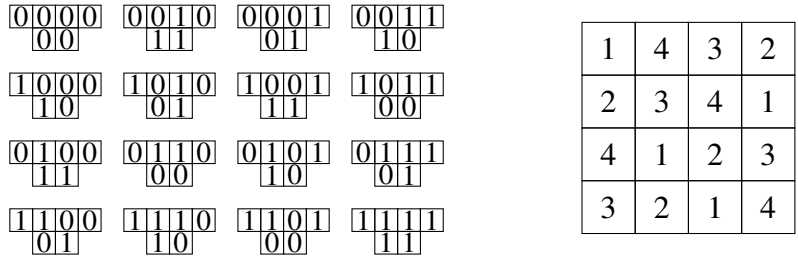
\begin{figure}[t]
    \centering
    \begin{subfigure}{.5\textwidth}
        \centering
        \begin{tikzpicture}
            [->,auto,node distance=1.5cm,
            empt node/.style={font=\sffamily,inner sep=0pt,minimum size=0pt},
            rect node/.style={rectangle,draw,font=\sffamily,minimum size=0.3cm, inner sep=0pt, outer sep=0pt}]
            
            \node [empt node] (e1) {};
            \node [rect node] (i111) [right=0.5cm of e1] {$0$};
            \node [rect node] (i112) [right=0cm of i111] {$0$};
            \node [rect node] (i113) [right=0cm of i112] {$0$};
            \node [rect node] (i114) [right=0cm of i113] {$0$};
            \node [rect node] (i115) [below=0cm of i112] {$0$};
            \node [rect node] (i116) [right=0cm of i115] {$0$};
            
            \node [rect node] (i121) [right=0.3cm of i114] {$0$};
            \node [rect node] (i122) [right=0cm of i121] {$0$};
            \node [rect node] (i123) [right=0cm of i122] {$1$};
            \node [rect node] (i124) [right=0cm of i123] {$0$};
            \node [rect node] (i125) [below=0cm of i122] {$1$};
            \node [rect node] (i126) [right=0cm of i125] {$1$};
            
            \node [rect node] (i131) [right=0.3cm of i124] {$0$};
            \node [rect node] (i132) [right=0cm of i131] {$0$};
            \node [rect node] (i133) [right=0cm of i132] {$0$};
            \node [rect node] (i134) [right=0cm of i133] {$1$};
            \node [rect node] (i135) [below=0cm of i132] {$0$};
            \node [rect node] (i136) [right=0cm of i135] {$1$};
            
            \node [rect node] (i141) [right=0.3cm of i134] {$0$};
            \node [rect node] (i142) [right=0cm of i141] {$0$};
            \node [rect node] (i143) [right=0cm of i142] {$1$};
            \node [rect node] (i144) [right=0cm of i143] {$1$};
            \node [rect node] (i145) [below=0cm of i142] {$1$};
            \node [rect node] (i146) [right=0cm of i145] {$0$};
            
            \node [rect node] (i211) [below=0.6cm of i111] {$1$};
            \node [rect node] (i212) [right=0cm of i211] {$0$};
            \node [rect node] (i213) [right=0cm of i212] {$0$};
            \node [rect node] (i214) [right=0cm of i213] {$0$};
            \node [rect node] (i215) [below=0cm of i212] {$1$};
            \node [rect node] (i216) [right=0cm of i215] {$0$};
            
            \node [rect node] (i221) [right=0.3cm of i214] {$1$};
            \node [rect node] (i222) [right=0cm of i221] {$0$};
            \node [rect node] (i223) [right=0cm of i222] {$1$};
            \node [rect node] (i224) [right=0cm of i223] {$0$};
            \node [rect node] (i225) [below=0cm of i222] {$0$};
            \node [rect node] (i226) [right=0cm of i225] {$1$};
            
            \node [rect node] (i231) [right=0.3cm of i224] {$1$};
            \node [rect node] (i232) [right=0cm of i231] {$0$};
            \node [rect node] (i233) [right=0cm of i232] {$0$};
            \node [rect node] (i234) [right=0cm of i233] {$1$};
            \node [rect node] (i235) [below=0cm of i232] {$1$};
            \node [rect node] (i236) [right=0cm of i235] {$1$};
            
            \node [rect node] (i241) [right=0.3cm of i234] {$1$};
            \node [rect node] (i242) [right=0cm of i241] {$0$};
            \node [rect node] (i243) [right=0cm of i242] {$1$};
            \node [rect node] (i244) [right=0cm of i243] {$1$};
            \node [rect node] (i245) [below=0cm of i242] {$0$};
            \node [rect node] (i246) [right=0cm of i245] {$0$};
            
            \node [rect node] (i311) [below=0.6cm of i211] {$0$};
            \node [rect node] (i312) [right=0cm of i311] {$1$};
            \node [rect node] (i313) [right=0cm of i312] {$0$};
            \node [rect node] (i314) [right=0cm of i313] {$0$};
            \node [rect node] (i315) [below=0cm of i312] {$1$};
            \node [rect node] (i316) [right=0cm of i315] {$1$};
            
            \node [rect node] (i321) [right=0.3cm of i314] {$0$};
            \node [rect node] (i322) [right=0cm of i321] {$1$};
            \node [rect node] (i323) [right=0cm of i322] {$1$};
            \node [rect node] (i324) [right=0cm of i323] {$0$};
            \node [rect node] (i325) [below=0cm of i322] {$0$};
            \node [rect node] (i326) [right=0cm of i325] {$0$};
            
            \node [rect node] (i331) [right=0.3cm of i324] {$0$};
            \node [rect node] (i332) [right=0cm of i331] {$1$};
            \node [rect node] (i333) [right=0cm of i332] {$0$};
            \node [rect node] (i334) [right=0cm of i333] {$1$};
            \node [rect node] (i335) [below=0cm of i332] {$1$};
            \node [rect node] (i336) [right=0cm of i335] {$0$};
            
            \node [rect node] (i341) [right=0.3cm of i334] {$0$};
            \node [rect node] (i342) [right=0cm of i341] {$1$};
            \node [rect node] (i343) [right=0cm of i342] {$1$};
            \node [rect node] (i344) [right=0cm of i343] {$1$};
            \node [rect node] (i345) [below=0cm of i342] {$0$};
            \node [rect node] (i346) [right=0cm of i345] {$1$};
            
            \node [rect node] (i411) [below=0.6cm of i311] {$1$};
            \node [rect node] (i412) [right=0cm of i411] {$1$};
            \node [rect node] (i413) [right=0cm of i412] {$0$};
            \node [rect node] (i414) [right=0cm of i413] {$0$};
            \node [rect node] (i415) [below=0cm of i412] {$0$};
            \node [rect node] (i416) [right=0cm of i415] {$1$};
            
            \node [rect node] (i421) [right=0.3cm of i414] {$1$};
            \node [rect node] (i422) [right=0cm of i421] {$1$};
            \node [rect node] (i423) [right=0cm of i422] {$1$};
            \node [rect node] (i424) [right=0cm of i423] {$0$};
            \node [rect node] (i425) [below=0cm of i422] {$1$};
            \node [rect node] (i426) [right=0cm of i425] {$0$};
            
            \node [rect node] (i431) [right=0.3cm of i424] {$1$};
            \node [rect node] (i432) [right=0cm of i431] {$1$};
            \node [rect node] (i433) [right=0cm of i432] {$0$};
            \node [rect node] (i434) [right=0cm of i433] {$1$};
            \node [rect node] (i435) [below=0cm of i432] {$0$};
            \node [rect node] (i436) [right=0cm of i435] {$0$};
            
            \node [rect node] (i441) [right=0.3cm of i434] {$1$};
            \node [rect node] (i442) [right=0cm of i441] {$1$};
            \node [rect node] (i443) [right=0cm of i442] {$1$};
            \node [rect node] (i444) [right=0cm of i443] {$1$};
            \node [rect node] (i445) [below=0cm of i442] {$1$};
            \node [rect node] (i446) [right=0cm of i445] {$1$};
            
        \end{tikzpicture}
    \end{subfigure}%
    \begin{subfigure}{.5\textwidth}
        \centering
        \begin{tikzpicture}
            [->,auto,node distance=1.5cm,
            empt node/.style={font=\sffamily,inner sep=0pt,minimum size=0pt},
            rect node/.style={rectangle,draw,font=\sffamily,minimum size=0.7cm, inner sep=0pt, outer sep=0pt}]
            \large
            
            \node [rect node] (s11) {$1$};
            \node [rect node] (s12) [right=0cm of s11] {$4$};
            \node [rect node] (s13) [right=0cm of s12] {$3$};
            \node [rect node] (s14) [right=0cm of s13] {$2$};
            
            \node [rect node] (s21) [below=0cm of s11] {$2$};
            \node [rect node] (s22) [right=0cm of s21] {$3$};
            \node [rect node] (s23) [right=0cm of s22] {$4$};
            \node [rect node] (s24) [right=0cm of s23] {$1$};
            
            \node [rect node] (s31) [below=0cm of s21] {$4$};
            \node [rect node] (s32) [right=0cm of s31] {$1$};
            \node [rect node] (s33) [right=0cm of s32] {$2$};
            \node [rect node] (s34) [right=0cm of s33] {$3$};
            
            \node [rect node] (s41) [below=0cm of s31] {$3$};
            \node [rect node] (s42) [right=0cm of s41] {$2$};
            \node [rect node] (s43) [right=0cm of s42] {$1$};
            \node [rect node] (s44) [right=0cm of s43] {$4$};

            \node [empt node] (e1) [below=0cm of s44] {\phantom{M}};
            
        \end{tikzpicture}
    \end{subfigure}%
    \caption{Latin square of order $4$ generated by rule $150$.}
    \label{fig:r150-sq}
\end{figure}

To fix the ideas discussed so far with an example, Figure~\ref{fig:r150-sq} shows the Cayley table associated to the NBCA $F: \F_2^4 \rightarrow \F_2^2$ defined by rule $150$. The bijective monotone mapping $\phi$ used here to encode pairs of bits $\phi(00) \mapsto 1$, $\phi(10) \mapsto 2$, $\phi(01) \mapsto 3$ and
$\phi(11) \mapsto 4$.

When $\Sigma = \F_q$ is the finite field of order $q$, the  result below proved in~\cite{mariot20} gives a characterization of linear Orthogonal CA (or linear OCA), i.e. pairs of linear bipermutive CA whose associated Latin squares are also orthogonal:
\begin{theorem}
\label{thm:lin-oca}
    Let $F,G:\F_q^{2(d-1)} \to \F_q^{d-1}$ be two linear bipermutive CA, and let $p_f,p_g \in \F_q[X]$ be the respective associated polynomials. Then, the corresponding Latin squares $C_F$ and $C_G$ of order $N=q^{d-1}$ are orthogonal if and only if $\gcd(p_f,p_g) = 1$, i.e. if and only if $p_f$ and $p_g$ are relatively prime.
\end{theorem}
The proof of Theorem~\ref{thm:lin-oca} stands on the observation that, in order for the Latin squares to be orthogonal, the square matrix obtained by stacking the transition matrices of $F$ and $G$ one on top of the other must be invertible. Crucially, the stacked matrix is the \emph{Sylvester matrix} of the two polynomials $p_f$ and $p_g$, and its determinant (also called the \emph{resultant}) vanishes if and only if $p_f$ and $p_g$ have at least one common factor.

Theorem~\ref{thm:lin-oca} gives a rather efficient method to check whether two linear CA are orthogonal: instead of generating exhaustively their associated Latin squares and checking if they are orthogonal, one can simply compute the greatest common divisor (e.g., by means of Euclid's algorithm) of the underlying polynomials.

We refer the reader to~\cite{mariot20} for further results leveraging on Theorem~\ref{thm:lin-oca}, including how to count the number of all linear OCA of a given diameter $d$, and how to construct a maximal family of Mutually Orthogonal Latin Squares based on linear CA (also called MOCA).

\section{Empirical Search of Self-Orthogonal CA}
\label{sec:emp}

A Latin square $L$ is called \emph{self-orthogonal} if it is orthogonal to its transpose $L^\top$~\cite{baker83}. This definition can be naturally adapted to the case of Latin squares generated by CA, yielding the concept of \emph{self-orthogonal CA}: 

\begin{definition}
\label{def:soca}
Let $F: \Sigma^{2(d-1)} \to \Sigma^{d-1}$ be a NBCA over an alphabet $\Sigma$ of size $q$ defined by a bipermutive local rule $f: \Sigma^d \to \Sigma_q$ of diameter $d \in \N$. Then, $F$ is a \emph{self-orthogonal CA} if the Latin square $S_F$ that it generates is self-orthogonal.
\end{definition}

Since we are mostly interested in applications to cryptography and coding theory, in what follows we focus on the case where the alphabet $\Sigma$ is the finite field $\F_q$, thus with its cardinality $q$ being the power of a prime number.

Clearly, one can check if a bipermutive NBCA is self-orthogonal by constructing the associated Latin square and superimposing it to its transpose, verifying that all ordered pairs in the Cartesian product of $(d-1)$-tuples occur exactly once. However, this procedure is exponential in the diameter of the automaton, since the NBCA $F$ has to be evaluated over all $q^{2(d-1)}$ input configurations. Therefore, \emph{we are interested in a more efficient characterization, which focuses on the properties of the underlying local rule}. This motivates the following problem, which we address in the remainder of the paper.
\begin{problem}
\label{pb:stat}
    Let $F: \F_q^{2(d-1)} \to \F_q^{d-1}$ be a NBCA of diameter $d \in \N$, defined by a bipermutive local rule $f: \F_q^d \to \F_q$. What conditions must the local rule $f$ satisfy for $F$ to be self-orthogonal?
\end{problem}

To get a general impression of this problem, let us start our investigation in an empirical manner. Consider the simplest case possible where the alphabet is binary (i.e., $\Sigma = \F_2 = \{0,1\}$) and the diameter is $d=3$, that is, the case of elementary CA. Here, up to complement and reversal of the inputs, there are only two bipermutive CA that give rise to Latin squares of order $2^{(3-1)} = 4$, namely those respectively defined by the linear local rules 90 and 150~\cite{mariot17}. By Theorem~\ref{thm:lin-oca}, these two Latin squares are orthogonal between each other, since the associated polynomials $1 + X^2$ and $1 + X + X^2$ do not have any factors in common. However, as depicted in Figure~\ref{fig:ex-90-150}, only rule 150 generates a self-orthogonal Latin square.
\begin{figure}[t]
    \centering
    \begin{subfigure}{.5\textwidth}
    \centering
    \begin{tikzpicture}
        [->,auto,node distance=1.5cm,
        empt node/.style={font=\sffamily,inner sep=0pt,minimum size=0pt},
        rect node/.style={rectangle,draw,font=\sffamily,minimum size=0.8cm, inner sep=0pt, outer sep=0pt}]
        \node [rect node] (s11) {$1,1$};
        \node [rect node] (s12) [right=0cm of s11] {$2,2$};
        \node [rect node] (s13) [right=0cm of s12] {$3,3$};
        \node [rect node] (s14) [right=0cm of s13] {$4,4$};
        \node [rect node] (s21) [below=0cm of s11] {$2,2$};
        \node [rect node] (s22) [right=0cm of s21] {$1,1$};
        \node [rect node] (s23) [right=0cm of s22] {$4,4$};
        \node [rect node] (s24) [right=0cm of s23] {$3,3$};
        \node [rect node] (s31) [below=0cm of s21] {$3,3$};
        \node [rect node] (s32) [right=0cm of s31] {$4,4$};
        \node [rect node] (s33) [right=0cm of s32] {$1,1$};
        \node [rect node] (s34) [right=0cm of s33] {$2,2$};
        \node [rect node] (s41) [below=0cm of s31] {$4,4$};
        \node [rect node] (s42) [right=0cm of s41] {$3,3$};
        \node [rect node] (s43) [right=0cm of s42] {$2,2$};
        \node [rect node] (s44) [right=0cm of s43] {$1,1$};
    \end{tikzpicture}
    \caption{Rule 90.}
    \end{subfigure}%
    \begin{subfigure}{.5\textwidth}
    \centering
    \begin{tikzpicture}
        [->,auto,node distance=1.5cm,
        empt node/.style={font=\sffamily,inner sep=0pt,minimum size=0pt},
        rect node/.style={rectangle,draw,font=\sffamily,minimum size=0.8cm, inner sep=0pt, outer sep=0pt}]
        \node [rect node] (s11) {$1,1$};
        \node [rect node] (s12) [right=0cm of s11] {$4,3$};
        \node [rect node] (s13) [right=0cm of s12] {$2,4$};
        \node [rect node] (s14) [right=0cm of s13] {$3,2$};
        \node [rect node] (s21) [below=0cm of s11] {$3,4$};
        \node [rect node] (s22) [right=0cm of s21] {$2,2$};
        \node [rect node] (s23) [right=0cm of s22] {$4,1$};
        \node [rect node] (s24) [right=0cm of s23] {$1,3$};
        \node [rect node] (s31) [below=0cm of s21] {$4,2$};
        \node [rect node] (s32) [right=0cm of s31] {$1,4$};
        \node [rect node] (s33) [right=0cm of s32] {$3,3$};
        \node [rect node] (s34) [right=0cm of s33] {$2,1$};
        \node [rect node] (s41) [below=0cm of s31] {$2,3$};
        \node [rect node] (s42) [right=0cm of s41] {$3,1$};
        \node [rect node] (s43) [right=0cm of s42] {$1,2$};
        \node [rect node] (s44) [right=0cm of s43] {$4,4$};
    \end{tikzpicture}
    \caption{Rule 150.}
    \end{subfigure}
    \caption{Self-superposition of rule 90 and 150. Only rule 150 is self-orthogonal, since all 16 pairs in the Cartesian product of $[4]\times [4]$ occur exactly once.}
    \label{fig:ex-90-150}
\end{figure}
Indeed, a closer look shows that the square generated by rule 90 is symmetric (i.e. it equals its transpose), hence it cannot be self-orthogonal.

The finding above prompted us to perform a more systematic exploration of self-orthogonal CA. Hence, we set out to enumerate in an exhaustive manner all binary bipermutive CA up to diameter $d=6$. For each local rule, we checked whether the corresponding bipermutive CA was self-orthogonal with the naive method mentioned earlier, i.e. by completely constructing the associated Latin square and its transpose, and checking whether all pairs in the Cartesian product occurred exactly once. As recalled in the previous section, there are $2^{2^{d-2}}$ bipermutive local rules of diameter $d$ over $\F_2$, since the generating function is determined by the central $d-2$ variables. Therefore, for the largest problem instance of $d=6$ we exhaustively checked $2^{16} = 65\,536$ rules for self-orthogonality. For each self-orthogonal Latin square found, we further checked whether it was generated by a nonlinear or linear rule, and in the latter case we computed the associated polynomial. Table~\ref{tab:emp-src} summarizes the results of the exhaustive search experiments, reporting for each diameter the corresponding number of bipermutive rule (\#BCA), the number of self-orthogonal CA found (\#SOCA), how many of them were linear (\#LIN) and finally the associated polynomials.
\begin{table}
\centering
\caption{Exhaustive enumeration of self-orthogonal CA up to diameter $d=6$.}
\scriptsize
\begin{tabular}{ccccc}
\toprule
$d$ & \#BCA   & \#SOCA & \#LIN & Polynomials \\
\toprule
3   & 4       & 2      & 2     & $1+X+X^2$ \\
\midrule
4   & 16      & 4      & 4     & $1+X+X^3$,
$1+X^2+X^3$\\
\midrule
5   & 256     & 8      & 8     & \makecell{
$1 + X + X^4$,
$1 + X^2 + X^4$ \\
$1 + X^3 + X^4$,
$1 + X + X^2 + X^3 + X^4$
} \\
\midrule
6   & 65\,336 & 16     & 16    & \makecell{
$1 + X + X^5$,
$1 + X^2 + X^5$ \\
$1 + X^3 + X^5$,
$1 + X^4 + X^5$ \\
$1 + X + X^2 + X^3 + X^5$ \\
$1 + X + X^2 + X^4 + X^5$ \\
$1 + X + X^3 + X^4 + X^5$ \\
$1 + X^2 + X^3 + X^4 + X^5$
} \\
\bottomrule
\end{tabular}
\label{tab:emp-src}
\end{table}

Two main facts stand out from the table: the first one is that, among those found, there are no self-orthogonal CA defined by nonlinear local rules. Indeed, the numbers in the columns \#SOCA and \#LIN coincide. The second fact is that the number of linear SOCA grows exponentially as the sequence of powers of $2$. Actually, one can notice a straightforward symmetry from these exhaustive search experiments: if a rule $f: \F_2^d \to \F_2$ generates a self-orthogonal CA, then so does its complement $f_C: \F_2^d \to \F_2$ defined as $f(x)_C = f(x) \oplus 1$ for all $x \in \F_2^d$. This is the reason why the polynomials reported in the last column amount to half of the number of linear self-orthogonal CA that have been found. The constant term in the ANF of the complemented local rule is simply neglected when considering the associated polynomial.

A third fact, which is a little less obvious from the table, is that for each considered diameter $d$ \emph{all irreducible polynomials of degree} $n=d-1$ appear in the last column. This seems to indicate that irreducibility of the associated polynomials is a sufficient condition for constructing a self-orthogonal CA. Notice that it cannot be necessary condition though, as---for example---the polynomial $1+X+X^5 = (1+X+X^2)(1 + X^2 + X^3)$ is not irreducible, and yet it appears in the list. We investigate this matter more in detail in the next section.

\section{Characterization of the Linear Case}
\label{sec:lin-ca}
Given the findings highlighted in the previous section concerning our exhaustive search experiments, we now narrow our attention to linear self-orthogonal CA, seeking a full algebraic characterization in the following subsections. We first explore them in the general case over the finite field $\F_q$, before specializing to the binary case, where we prove a few more results that explain our findings.

\subsection{General Results for the Alphabet $\Sigma = \F_q$}
\label{subsec:gen}
Given a linear NBCA $F: \F_q^{2(d-1)} \to \F_q^{d-1}$, let us start by analyzing more in depth what does it mean to compute the transpose of $F$. Following the notation laid out in Section~\ref{subsec:ls}, transposing the Cayley table of the CA means that we first swap the left half and the right half of its input configuration, and then we apply the global map $F$ of the CA. Consequently, the transpose CA $F^\top: \F_q^{2(d-1)} \to \F_q^{d-1}$ is defined for all $x, y \in \F_q^{d-1}$ as:
\begin{equation}
\label{eq:trans-ca}
F^\top(x||y) = F(y||x) \enspace .
\end{equation}
In particular, given $x \in \F_q^n$ with $n \in \N$ even, the swap operation can be formalized as the linear transformation $P_s(x) = M_s \cdot x^\top$, where $M_s$ is the $n \times n$ \emph{permutation matrix} defined as:
\begin{equation}
\label{eq:perm-mat}
M_s = 
\begin{pmatrix}
0 & 0 & \ldots & 0 & 1 & 0 & \ldots & 0 & 0 \\
0 & 0 & \ldots & 0 & 0 & 1 & \ldots & 0 & 0 \\
\vdots & \vdots & \ddots & \vdots & \vdots & \vdots & \ddots & \vdots & \vdots \\
0 & 0 & \ldots & 0 & 0 & 0 & \ldots & 0 & 1 \\
1 & 0 & \ldots & 0 & 0 & 0 & \ldots & 0 & 0 \\
0 & 1 & \ldots & 0 & 0 & 0 & \ldots & 0 & 0 \\
\vdots & \vdots & \ddots & \vdots & \vdots & \vdots & \ddots & \vdots & \vdots \\
0 & 0 & \ldots & 0 & 1 & 0 & \ldots & 0 & 0 \\
\end{pmatrix}
\end{equation}
Therefore, in the case of a linear CA $F: \F_q^{2(d-1)}$, the global map of its transpose $F^\top$ can also be expressed as a linear transformation $F^\top(x) = M_{F^\top} \cdot x^\top$, where $M_{F^\top}$ is the $(d-1) \times 2(d-1)$ matrix obtained by multiplying the transition matrix $M_F$ of $F$ with the permutation matrix $M_s$ defined in Eq.~\eqref{eq:perm-mat}: 
\begin{equation}
\tiny
\label{eq:mult-mat}
M_{F^\top} = M_F \cdot M_{s} =
\begin{pmatrix}
        a_1    & \ldots & \ldots & a_d & 0 & \ldots & \ldots & \ldots & \ldots & 0 \\
        0      & a_1    & \ldots  & \ldots & a_d & 0 & \ldots & \ldots & \ldots & 0 \\
        \vdots & \vdots & \vdots & \ddots  & \vdots & \vdots & \vdots & \ddots & \vdots & \vdots \\
        0 & \ldots & \ldots & \ldots & \ldots & 0 & a_1 & \ldots & \ldots & a_d \\
\end{pmatrix}
\cdot
\begin{pmatrix}
0 & 0 & \ldots & 0 & 1 & 0 & \ldots & 0 & 0 \\
0 & 0 & \ldots & 0 & 0 & 1 & \ldots & 0 & 0 \\
\vdots & \vdots & \ddots & \vdots & \vdots & \vdots & \ddots & \vdots & \vdots \\
0 & 0 & \ldots & 0 & 0 & 0 & \ldots & 0 & 1 \\
1 & 0 & \ldots & 0 & 0 & 0 & \ldots & 0 & 0 \\
0 & 1 & \ldots & 0 & 0 & 0 & \ldots & 0 & 0 \\
\vdots & \vdots & \ddots & \vdots & \vdots & \vdots & \ddots & \vdots & \vdots \\
0 & 0 & \ldots & 0 & 1 & 0 & \ldots & 0 & 0 \\
\end{pmatrix}
\end{equation}
In particular, Eq~\eqref{eq:mult-mat} can be rewritten as:

\begin{equation}
\label{eq:trans-mat}
M_{F^\top} = M_F \cdot M_{s} =
\begin{pmatrix}
        a_d & 0 & \ldots & \ldots & 0 & a_1 & \ldots & \ldots & a_{d-1} \\
        a_{d-1} & a_d & \ldots & \ldots & 0 & 0 & a_1 & \ldots & a_{d-2} \\
        \vdots & \vdots & \ddots & \ddots & \vdots & \ddots & \ddots & \ddots & \vdots \\
        a_2 & a_3 & \ldots & \ldots & a_d & 0 & 0 & \ldots & a_1
\end{pmatrix} \enspace .
\end{equation}
Equivalently, the transition matrix of the transpose CA of $F^\top$ is obtained by swapping the left and the right half of the transition matrix of $F$.

Consider now the $2(d-1) \times 2(d-1)$ square matrix $M_{F,F^\top}$ obtained by stacking $M_F$ on top of $M_F^\top$:

\begin{equation}
    \label{eq:stack-mat}
    M_{F,F^\top} = 
    \begin{pmatrix}
        M_F \\
        M_{F^\top}
    \end{pmatrix}
    =
    \begin{pmatrix}
        a_1 & \ldots & \ldots & a_{d} & 0 & 0 & \ldots & \ldots & 0\\
        0 & a_1 & \ldots & \ldots & a_d & 0 & \ldots &\ldots & 0 \\
        \vdots & \vdots & \ddots & \ddots & \vdots & \ddots & \ddots & \ddots & \vdots \\
        0 & 0 & \ldots & \ldots & a_1 & \ldots & \ldots & \ldots & a_d\\
        a_d & 0 & \ldots & \ldots & 0 & a_1 & \ldots & \ldots & a_{d-1} \\
        a_{d-1} & a_d & \ldots & \ldots & 0 & 0 & a_1 & \ldots & a_{d-2} \\
        \vdots & \vdots & \ddots & \ddots & \vdots & \ddots & \ddots & \ddots & \vdots \\
        a_2 & a_3 & \ldots & \ldots & a_d & 0 & 0 & \ldots & a_1
\end{pmatrix} \enspace .
\end{equation}
Then, we can prove the following necessary and sufficient condition for the self-orthogonality of $F$:
\begin{lemma}
\label{lm:main}
The linear bipermutive CA $F: \F_q^{2(d-1)} \to \F_q^{d-1}$ is self-orthogonal if and only if the $2(d-1) \times 2(d-1)$ square matrix defined in Eq.~\eqref{eq:stack-mat} is invertible.
\begin{proof}
    The proof is analogous to the one of Theorem~\eqref{thm:lin-oca} reported in~\cite{mariot20}. In particular, let $M_F$ be the transition matrix of $F$, and let $F^\top: \F_q^{2(d-1)} \to \F_q^{d-1}$ be the transpose CA with $M_{F^\top}$ being its associated matrix. Then, $F$ is self-orthogonal if and only if the superposed mapping $\mathcal{H}: \F_q^{d-1}  \times \F_q^{d-1} \to \F_q^{d-1} \times \F_q^{d-1}$ defined for all $(x,y) \in \F_q^{2(d-1)}  \times \F_q^{2(d-1)}$ as:
    \begin{equation}
        \label{eq:map-h}
        \mathcal{H}(x,y) = (F(x||y), F^\top(x||y)) = (w, z)
    \end{equation}
    is bijective. The bijectivity of $\mathcal{H}$ amounts to the fact that the following system of $2(d-1)$ linear equations has exactly one solution for each possible assignment of the $2(d-1)$ variables:
    \begin{equation}
        \label{eq:syst}
        \begin{cases}
            F(x||y) = M_F(x||y)^\top = w \\
            F^\top(x||y) = M_{F^\top} (x||y)^\top = z
        \end{cases} \enspace .
    \end{equation}
    In other words, the first half of the output $w$ of $\mathcal{H}$ corresponds to the application of the CA $F$ on the input obtained by concatenating $x$ with $y$. Symmetrically, the second half $z$ is obtained by computing the transpose CA $F^\top$ over the same concatenation. Finally, remark that the matrix associated to the system above is exactly the stacked matrix of Eq.~\eqref{eq:stack-mat}. Hence, $\mathcal{H}$ is bijective if and only if the matrix~\eqref{eq:stack-mat} is invertible.
\end{proof}
\end{lemma}

Differently from the case where we consider the orthogonality of the Latin squares generated by two distinct linear bipermutive CA, in our setting the square matrix defined in Eq.~\eqref{eq:stack-mat} is not of the Sylvester type. However, one can notice that this matrix is \emph{circulant}. In particular, each row is the \emph{cyclic right shift} of the row above. Formally, denoting $M_{F,F^\top} = (c_{ij})$ for all $i,j \in \{0,\cdots, n-1\}$, we have that $c_{i+1,j+1} = c_{i,j}$, where all indices are computed modulo $n=2(d-1)$.

Fortunately, a great deal is known about circulant matrices, especially concerning their invertibility properties (see e.g. Section 13.2.4 of~\cite{mullen13} for a summary of the main results). In particular, the set $R_c$ of all $n \times n$ circulant matrices over a field is a ring, and thus it is closed under matrix addition and multiplication. Interestingly, circulant matrices are isomorphic to polynomials, as the next result shows:
\begin{theorem}[\cite{lidl94}]
\label{thm:isom}
    Let $R_C$ be the ring of $n\times n$ circulant matrices over the finite field $\F_q$, and let $R_P = \F_q[X] / (X^n - 1)$ be the quotient polynomial ring over $\F_q$ modulo $(X^n-1)$. Given a matrix $M \in R_C$ whose first row is the vector $(c_1,\ldots, c_n)$, the following mapping:
    \begin{equation}
        \label{eq:isom}
        \Phi : (c_1,\ldots,c_n) \mapsto c(X) = c_1 + c_2X + \ldots + c_nX^{n-1} \mod{X^n - 1}
    \end{equation}
    is an isomorphism of rings between $R_C$ and $R_P$. 
\end{theorem}
Notice that the map $\Phi$ of Eq.~\eqref{eq:isom} is exactly the same that we used to associate the polynomial $p_f(X)$ to a linear local rule $f$ in Section~\ref{subsec:ca}. From the perspective of the circulant matrix, performing a right cyclic shift of the coefficients corresponds to multiplying the polynomial $c(X)$ by the monomial $X$, and then reducing the result modulo $X^n - 1$. Thus, in general, the $i$-th row of a circulant matrix is obtained by taking the coefficients of the following polynomial:
\begin{equation}
    \label{eq:poly-i}
    c_i(X) = X^i c(X) \mod{X^n - 1} \enspace . 
\end{equation}

From a practical point of view, Theorem~\ref{thm:isom} states that one can check if a circulant matrix $M \in R_C$ is invertible by checking if the corresponding polynomial $c(X) =\Phi(M)$ is a \emph{unit} in the quotient ring $\F_q[X] / (X^n - 1)$, that is if it has a multiplicative inverse $c^{-1}(X)$ such that $c(X)\cdot c^{-1}(X) = 1$. However, this is equivalent to computing the greatest common divisor of $c(X)$ and $X^n - 1$. In particular, $c(X)$ is invertible if and only if $\gcd(c(X), X^n-1) = 1$. We have thus proved the following characterization result:
\begin{theorem}
\label{thm:main}
Let $F: \F_q^{2(d-1)} \to \F_q^{d-1}$ be a NBCA of size $n=2(d-1)$ with a linear bipermutive rule $f: \F_q^d \to \F_q$ of diameter $d$ and associated polynomial $p_f(X) \in \F_q[X]$. Then, $F$ is self-orthogonal if and only if $\gcd(p_f(X), X^n-1) = 1$.
\end{theorem}

\subsection{The Special Case $\Sigma = \F_2$}
\label{subsec:bin}
Theorem~\ref{thm:main} gives an elegant way to check if a linear bipermutive CA over any finite field is self-orthogonal: instead of generating the full Latin square and the related transpose, one can simply apply Euclid's algorithm to check if the greatest common divisor of the associated polynomial with $X^n-1$ is 1. This gives us a result analogous to the generic case addressed in~\cite{mariot20}, since the orthogonality of two CA is again related to the coprimality of two polynomials.

In this section, we consider the specific variant of Problem ~\ref{pb:stat} when the underlying alphabet is the binary field $\F_2$, due to its particular interest for practical applications in cryptography and coding theory. As we are working in characteristic 2, we have $X^n - 1 = X^{2(d-1)} + 1 = (X^{d-1} + 1)^2$. If $p_f(X)$ has a non-trivial factor in common with $X^n - 1$, then it must also have a non-trivial factor in common with $X^{d-1} + 1$. Hence, in the particular case of $\F_2$, the condition provided in Theorem~\ref{thm:main} is equivalent to $\gcd(p_f(X), X^{d-1} + 1) = 1$, and Theorem~\ref{thm:main} can be restated as:
\begin{corollary}
\label{cor:mainF2}
Let $F: \F_2^{2(d-1)} \to \F_2^{d-1}$ be a NBCA of size $n=2(d-1)$ with a linear bipermutive rule $f: \F_2^d \to \F_2$ of diameter $d$ and associated polynomial $p_f(X) \in \F_2[X]$. Then, $F$ is self-orthogonal if and only if $\gcd(p_f(X), X^{d-1}+1) = 1$.
\end{corollary}

We now prove that, in $\F_2$, irreducibility is a sufficient condition for self-orthogonality:
\begin{lemma}
\label{lm:irr}
Let $F: \F_2^{2(d-1)} \to \F_2^{d-1}$ be a NBCA of size $n=2(d-1)$ defined by a linear bipermutive rule $f: \F_2^d \to \F_2$ of diameter $d > 2$, whose associated polynomial is $p_f(X) \in \F_2[X]$. If $p_f(X)$ is irreducible, the NBCA $F$ is self-orthogonal.
\begin{proof}
By Corollary~\ref{cor:mainF2}, we have to show that $\gcd(p_f(X), X^{d-1} + 1) = 1$. Since $X = 1$ is a root of $X^{d-1} + 1$, we can rewrite it as
\begin{equation}
\label{eq:dec}
X^{d-1} + 1 = (X + 1) \, q(X) \enspace ,   
\end{equation}
for some $q \in \F_2[X]$ with degree $d - 2$. However, $f$ is an irreducible polynomial with degree $d-1$, thus it cannot be neither $X + 1$ (as $d > 2$) nor $q(X)$ in the right-hand side of Eq.~\eqref{eq:dec}. Therefore $\gcd(p_f(X), X^{d-1} + 1) = 1$, and $F$ is self-orthogonal.
\end{proof}
\end{lemma}
Being a sufficient condition, Lemma~\ref{lm:irr} explains why, in the results of the exhaustive search reported in Section~\ref{sec:emp}, all irreducible polynomials induce self-orthogonal linear CA.

We now consider another particular case, i.e. when the size of the stacked matrix in Eq.~\eqref{eq:stack-mat} is a power of $2$.
\begin{lemma}
\label{lm:pow-2}
Let $d = 2^t + 1$ for some $t$. Then, a linear bipermutive CA $F:\F_2^{2(d-1)} \to \F_2^{d-1}$ defined by the local rule $f: \F_2^d \to \F_2$ with associated polynomial $p_f(X)$ is self-orthogonal if and only if $p_f(1) \neq 0$.
\begin{proof}
Since $\F_2$ is a field of characteristic 2 and $d - 1$ is a power of 2, we have that $X^{d-1} + 1 = X^{2^t} + 1 = (X + 1)^{d-1}$.
This easily follows by induction and by the fact that $X^2 + 1 = (X + 1)^2$ in characteristic 2.
Hence, by Corollary~\ref{cor:mainF2} $F$ is self-orthogonal if and only if $\gcd(p_f(X), (X+1)^{d-1}) = 1$. Any nontrivial common divisor of $p_f(X)$ and $(X+1)^{d-1}$ must be of the form $(X+1)^k$ for some $k < d$. Therefore, $p_f(X)$ and $(X+1)^{d-1}$ do not share common factors if and only if $p_f(1) \neq 0$.
\end{proof}
\end{lemma}

The condition $p_f(1) \neq 0$ of Lemma~\ref{lm:pow-2} can be further refined in a very simple procedure: given a polynomial $f(X) \in \F_2[X]$, one needs to check the \emph{parity} of its coefficients. Indeed, setting $X=1$ means that we compute the XOR of all the coefficients of $f(X)$. Therefore, one has that $f(1) = 0$ if the number of coefficients of $f$ set to $1$ is even, and $f(1) = 1$ if it is odd.

Lemma~\ref{lm:pow-2} partly explains the numbers of linear self-orthogonal CA observed in the exhaustive search experiments of Section~\ref{sec:emp}: in particular, the cases $d=3$ and $d=5$ yield square matrices whose size is a power of $2$ (respectively, $2$ and $4$). Then, since we are considering linear bipermutive CA, we know that the leftmost and rightmost coefficients of the associated polynomials are always not null. Hence, the condition $f(1) \neq 0$ is determined only by the parity of the central $d-2$ coefficients. The number of combinations with an odd parity is exactly half of all $2^{d-2}$ combinations of the central coefficients, i.e. $2^{d-3}$. We have thus obtained the following counting result:

\begin{lemma}
\label{lm:count}
Let $d = 2^t + 1$ for some $t$. Then, the number of self-orthogonal CA defined by linear bipermutive rules of the form $f: \F_2^d \to \F_2$ is $2^{d-3}$.
\end{lemma}

However, Lemma~\ref{lm:count} does not account for the fact that the number of linear self-orthogonal CA seems also to follow the sequence $2^{d-3}$ when $d-1$ is not a power of $2$. For this reason, we used Theorem~\ref{thm:main} to extend our exhaustive search experiments, by generating all linear self-orthogonal CA up to diameter $d=16$. As it can be seen from Table~\ref{tab:lin-soca}, the pattern of the sequence $2^{d-3}$ already breaks from diameter $d=7$, which is the smallest case that we have not checked exhaustively in Section~\ref{sec:emp}.
\begin{table}[t]
    \centering
    \scriptsize
    \begin{tabular}{cccccccccccccccc}
    \toprule
       $d$ & 3 & 4 & 5 & 6 & 7 & 8 & 9 & 10 & 11 & 12 & 13 & 14 & 15 & 16 \\
    \midrule
       \#LIN & 1 & 2 & 4 & 8 & 12 & 24 & 64 & 94 & 240 & 512 & 768 & 2048 & 3136 & 5062 \\
    \bottomrule
    \end{tabular}
    \caption{Number of linear self-orthogonal CA up to diameter}
    \label{tab:lin-soca}
\end{table}
The sequence does not seem to be already known in the Online Encyclopedia of Integer Sequences (OEIS). We leave the point of deriving a general recurrence equation for the number of linear self-orthogonal CA as an open problem for future research.

As a final remark, notice that Lemmas~\ref{lm:irr} and~\ref{lm:pow-2} hold over any finite field of characteristic 2. Therefore, they are directly applicable to self-orthogonal CA defined over any finite field $\F_{2^k}$ as an alphabet.

\section{Conclusions and Open Problems}
\label{sec:outro}
In this paper, we addressed the problem of determining when a bipermutive CA is self-orthogonal, i.e. when it generates a Latin square that is orthogonal to its transpose. We started with an empirical approach by enumerating all self-orthogonal binary CA up to diameter $d=6$, noticing that only some linear CA are self-orthogonal. We thus tackled the linear case more systematically, proving a characterization that involves computing the greatest common divisor of the polynomial associated to the local rule and the polynomial $X^{2(d-1)}-1$. Moreover, we looked at the binary case more in detail, proving that irreducibility is indeed a sufficient condition for self-orthogonality, as observed in our exhaustive search experiments. On the other hand, we showed a simpler characterization of linear self-orthogonal CA over the binary alphabet when the size of the matrix is a power of 2. In this case, the question whether the CA is self-orthogonal reduces to checking the parity of the central coefficients of the local rule.

To conclude, we take a step back on our results and focus on the main condition for self-orthogonality, i.e. the invertibility of the corresponding circulant matrix. Circulant matrices are already well known in the CA literature: indeed, the transition matrix of a linear PBCA is a circulant matrix, as noted for example by Bini et al.~\cite{bini98}. However, this connection remained relatively incospicuous in the literature, since most of the related works investigate the invertibility of \emph{infinite} linear CA over a ring, and thus resort to the machinery of formal power series~\cite{Ito83,manzini98}. Thus, our work shows that \emph{studying the self-orthogonality of linear bipermutive NBCA is equivalent to determining whether they are invertible as PBCA}. This remark gives an interesting link between the combinatorial perspective on bipermutive CA and their dynamical properties.

For future work, there are several interesting directions to explore on the subject. In particular, the most evident open problem is to determine a counting formula for the number of linear self-orthogonal CA in the general case, which from our experiments seems to follow the sequence of the powers of 2. Further, it would be interesting to assess, even experimentally, if \emph{nonlinear} self-orthogonal CA exist. Possibly, the spaces of rules of diameter up to $d=6$ considered in this paper is too limited to grant the existence of nonlinear self-orthogonal CA. Finally, one general research direction to explore is whether CA can be used to construct families of \emph{mutually self-orthogonal Latin squares}~\cite{baker83}. In particular, these objects can be used to obtain resolvable BIBDs, from which anonymous secret sharing schemes~\cite{blundo97} and Mutually Unbiased Bases for quantum error-correcting codes~\cite{kumar22} can be further constructed.

\bibliographystyle{abbrv}
\bibliography{bibliography}

\end{document}